\documentclass{article}
\usepackage{graphicx}  
\usepackage{amsmath}   
\usepackage{amssymb}   
\usepackage{bm} 
\usepackage{dcolumn}
\usepackage{color}
\usepackage{mathrsfs}
\usepackage{amsfonts}
\usepackage{varioref}
\RequirePackage[colorlinks,citecolor=blue,urlcolor=magenta,linkcolor=blue]{hyperref}
\def\be{\begin{equation}}
\def\ee{\end{equation}}
\labelformat{section}{Section #1} 
\labelformat{subsection}{Section #1} 
\labelformat{subsubsection}{Section #1}
\labelformat{subsubsubsection}{Section #1}
\labelformat{equation}{Eq.~(#1)} 
\labelformat{figure}{Fig.~#1} 
\labelformat{subfigure}{Fig.~\thefigure#1} 
\labelformat{table}{Tab.~#1} 
\labelformat{appendix}{Appendix #1}

\begin{document}

\title{\bf The maximum turnaround radius  of non-spherical cosmic structures}
\author{Sourav Bhattacharya$^1$\footnote{sbhatta@iitrpr.ac.in}~ and  Theodore N Tomaras$^2$\footnote{tomaras@physics.uoc.gr}\\
$^{1}$\small{Department of Physics, Indian Institute of Technology Ropar, Rupnagar, Punjab 140 001, India}\\
$^{2}$\small{ITCP and Department of Physics, University of Crete, Heraklion 700 13, Greece} }
\maketitle
%
\begin{abstract}
\noindent
Three simple idealised models are studied in order to develop some intuition about the leading order effect of non-sphericity on the maximum turnaround size $R_{\rm TA,max}$ of large scale bound cosmic structures.  Two of them describe intrinsically axisymmetric static mass distributions whereas the other is the Kerr-de Sitter metric where the axisymmetry  is generated   due to the rotation of the structure. In all the cases the fractional change $\delta R_{\rm TA,max}(\theta)/R^{(0)}_{\rm TA,max}$ of $R_{\rm TA,max}$ of a given structure, compared to a spherical one with the same mass $M$, depends on the polar angle $\theta$ and is proportional to the product of the relevant eccentricity parameter, times the square of a small quantity. This quantity in the static examples is the ratio of two characteristic length scales, while in  the spinning case it is the ratio $v_{\rm out}/c$ of the azimuthal speed of the outmost members of the structure, over the speed of light. Furthermore, the angular average $\langle \delta R_{\rm TA,max}(\theta)/R^{(0)}_{\rm TA,max}\rangle$ is zero in the two static cases, while it is negative and proportional to ${\cal O}(v^2_{\rm out}/c^2)$ for the Kerr-de Sitter. Thus, $\delta R_{\rm TA,max}(\theta)/R^{(0)}_{\rm TA,max}$ for an axisymmetric structure is very small for practically any value of the eccentricity parameter. We speculate about some possible further implications of our result on the maximum turn around radius of realistic cosmic structures.  
\end{abstract}
\vskip .3cm

\noindent
{\bf keywords :} Large scale structures, non-sphericity, maximum turnaround radius
\bigskip
\section{Introduction}\label{s1}

The concordance $\Lambda{\rm CDM}$ model is simple and, up to now, very successful in explaining the observational data~\cite{Weinberg:2008zzc, Lapuente}. 

However, it is still not entirely satisfactory for a variety of well-known reasons. At the theoretical front, there is yet no generally accepted fundamental-physics driven candidate for the nature of vacuum energy, while it requires severe fine tuning when interpreted as a cosmological constant \cite{Martin2012}. Furthermore, $\Lambda$CDM does not provide any insight towards a natural explanation of the ``cosmic coincidence problem", i.e., the fact that the current observed values of the dark energy and the cold dark matter energy densities are so close to each other. Perhaps even more importantly, at the observational front the increase in accuracy of individual cosmological datasets is starting to reveal tensions  (see e.g. \cite{cepheids2,  PlanckCosmology2018,   divalentino2019, Handley:2019tkm} and references therein) 
and have lead to frequent critical revisits of the strength of the evidence for the specific values of the cosmological parameters 
\cite{Rocky2006,  sarkar2016}, also references therein.

Motivated by the above issues, the community has plunged into intense research on alternatives of the $\Lambda{\rm CDM}$ model or Einstein's theory of gravitation, see e.g.~\cite{CliftonEtAl2012,  Cataneo:2018mil} for exhaustive reviews. To distinguish all these alternative gravity models from each other {\it and} from the $\Lambda{\rm CDM}$, we require suitable cosmological or astrophysical observable quantities.

In recent years, considerable attention has been given to the properties of cosmic structures on the largest scales as a means to locally probe cosmology and alternative theories of gravity (e.g., \cite{TC2008, PT14, PTT2014, splashback, Dragan,boundary1, Susmita2018}).
The turnaround radius in particular, i.e. the scale on which a cosmic structure detaches from the Hubble flow, has been the focus on many such studies \cite{PTT2014, TPT2015, TPT2016, LeeLi2017, Bhattacharya:2016vur,Faraoni2018, Dia2019, Santa2019, Lopes2019, Wong19, Ali:2017ofu}.

The turnaround radius, which can be measured kinematically in any galaxy cluster, as the boundary between the cluster and the expanding Universe, is sensitive to the presence of a cosmological constant $\Lambda$. Once the repulsive effect of $\Lambda$ becomes dominant over the gravitational self-attraction of matter, it halts structure growth (\cite{PT14, TPT2015, Busha1,Busha2}). As a consequence, in an ever expanding Universe with $\Lambda$ not all overdensities are destined to detach from the Hubble flow (e.g. \cite{PF05}). In $\Lambda$CDM in particular, the {\it maximum turnaround radius} $R^{(0)}_{\rm TA,max}$, the maximum possible size  of a spherical structure of mass $M$, i.e. the distance from its center at which the two forces on a static test particle balance each other, has a hard upper bound $(3M G/\Lambda c^2)^{1/3}$ \cite{PT14,Bhattacharya:2016vur}, or equivalently, the turnaround density $\rho_{\rm TA}$ has a hard lower bound $\rho^{(0)}_{\rm TA,min}= 2\rho_\Lambda = 2(\Lambda c^2/8\pi G)$, as predicted also by the spherical collapse model \cite{PT14}.

Remarkably, this theoretical prediction of concordance $\Lambda$CDM is in very good agreement, lying close from above to the observed sizes of superclusters with masses $M\gtrsim 10^{15}M_{\odot}$, assumed roughly spherically symmetric and isolated \cite{PT14}, with the fractional deviation from the predicted maximum size being estimated to about $10-20\%$. This means firstly that the $\Lambda{\rm CDM}$ is consistent with the observed bound cosmic structures, and secondly, that the size at turnaround of cosmic structures is a useful additional observable, which can be applied to nearby structures to obtain a {\it local} measurement of the cosmological constant in $\Lambda$CDM, or more generally to constrain alternative gravity models.

We refer our reader to \cite{Einasto:2016rhe}-\cite{Santa:2019snv} for various computations, data and parameter space analysis and other applications pertaining the $R_{\rm TA, max }$. The computations yield the same result no matter whether one uses a static or time dependent cosmological geometry. We further refer our reader to \cite{Bhattacharya:2017yix} for a more detailed review on this.

All the above theoretical studies were restricted to isolated spherically symmetric large scale cosmic structures. Since the turnaround radius refers to the size of a structure as it detaches from an expanding spherically symmetric environment, and is typically much bigger than the Schwarzschild radius of the structure, this is expected to be a reasonable approximation, and has to a large extent been confirmed as such by the level of agreement of observations with the results of the theoretical analyses based on the spherical collapse model, see e.g. \cite{PT14}. 
Nevertheless, it is still important to understand quantitatively and directly the effect of non-sphericity on the turnaround radius or the maximum turnaround radius $R_{\rm TA, max}$ of a cosmic structure of a given mass. In particular, this question seems very appropriate given the generic irregular form of the halos in the Cosmos and in numerical simulations.

In \cite{Barrow1, Barrow:1984zz} we find early discussions on gravitational collapse and virialisation with or without a $\Lambda$, in spatially inhomogeneous or anisotropic cosmological backgrounds. Recently the question was studied in \cite{Giusti:2019uez, Giusti:2019lha, Kunz:2019}, with approaches varying in generality, rigor and approximations. 

Encouraged by the success of the analytical treatment based on spherical symmetry and on the spherical collapse model in fitting the sizes of large superclusters, we will use here the same method to study analytically deviations from spherical symmetry in the context of $\Lambda$CDM. Specifically, we shall focus on axially symmetric structures using simple models as a first step away from spherical symmetry. We shall define appropriately the maximum turnaround surface of such a structure and study its dependence on the relevant eccentricity parameter and length scales. Perhaps this may be thought of as the simplest approach to address, in particular, the weak gravitational field of our nearby Corona-Borealis supercluster, which seems to be a binary connected via a filament, e.g. \cite{Pillastrini:2015oqa}. However, our purpose here is mainly to use this approach as a way to obtain intuition about the effects of eccentricity on the turnaround sizes of realistic structures, even though they are not axially  symmetric. In the same way that the study of isolated spherical structures and the use of the spherical collapse model gave us important information about the sizes of actual structures in the Universe. 

Note that an axisymmetric spacetime geometry cannot be unique like the spherical one. The only exception to this should be an isolated Kerr black hole, whose uniqueness is known. Thus in order to study the leading effect of non-sphericity on the maximum turn around radius of large scale structures, we must consider some suitable models. Keeping also in mind the axisymmetry of a structure can be either intrinsic or due to its rotation, we shall be interested in three different models as follows. 

For orientation, we start in \ref{newton} with a pedestrian Newtonian analysis of a homogeneous spheroidal structure \cite{pohanka:2011} in the presence of a cosmological constant $\Lambda$. In \ref{non-sph} we extend the analysis to the framework of the general-relativistic backgrounds of a static axisymmetric structure \cite{Glampedakis:2005cf, Johannsen:2010xs, Loeb:2013lfa}, while in \ref{kds} we model an  axisymmetric structure with the Kerr-de Sitter spacetime, in analogy to the Schwarzschild-de Sitter metric that was used in the study of the spherically symmetric case. For the first two cases, the axisymmetry is intrinsic, whereas for the Kerr-de Sitter, the axisymmetry is generated by the rotation of the structure. In the final \ref{conc} we summarise our results and comment on their relevance to actual  structures in the Universe. 

We shall use mostly positive signature of the metric and henceforth will set $c=1=G$.


\vskip -1cm

\section{Turnaround radius of a spheroidal structure in de Sitter}\label{newton}
We start with the Newtonian approximate treatment of the maximum turnaround radius in the gravitational field of a homogeneous oblate spheroid with semi-axes $\alpha$ and $\beta$ ($\alpha\geq \beta$) and total mass $M$ in a de Sitter background with cosmological constant $\Lambda=3H_0^2$, where $H_0^{-1} \sim 1.3\times 10^{10}\, {\rm ly}$ is approximately the inverse of the Hubble parameter today. This should be thought of as a simple toy model, a first step in the investigation of the turnaround size of large cosmic structures away from spherical symmetry, which can be analysed in detail and, as we shall see, offers useful intuition for the study of more realistic cases.

In the $(v,\xi,\psi)$ ellipsoid coordinate system of the oblate spheroid, the Newtonian potential outside it (i.e. for $v>1$) is the harmonic function \cite{pohanka:2011}
\be
V_N=-\frac{M}{\alpha\epsilon}\left(\cot^{-1}\sigma + \frac{1}{2} \left((3\sigma^2+1)\cot^{-1}\sigma - 3\sigma\right) P_2(\cos\xi)\right)
\label{VN}
\ee
with $\epsilon=\sqrt{1-\beta^2/\alpha^2}$ the ``eccentricity" of the ellipsoid, $\sigma=\sigma(v)=\sqrt{1-\epsilon^2}\,v/\epsilon$ and $P_2(x) = (3x^2-1)/2$ is the third Legendre polynomial. 

The position vector in the coordinate system $(v,\xi,\psi)$ is
$$
{\bf r} = \alpha \left(\sqrt{(1-\epsilon^2)v^2+\epsilon^2} \sin\xi\left(\hat{\bf i} \cos\psi + \hat{\bf j} \sin\psi\right) +\hat{\bf k} \sqrt{1-\epsilon^2} v\cos\xi\right)
$$
and, consequently, the relation of the ellipsoidal coordinates to the spherical polar ones $(r,\theta,\phi)$ is
\be
v=\frac{s(r,\theta)}{\alpha \sqrt{1-\epsilon^2}}\,, \;\;\;\; \cos\xi=\frac{r\cos\theta}{s(r,\theta)} \,, \;\;\;\; \psi=\phi
\label{conversion}
\ee
with
\be
s^2=\frac{1}{2} \left(r^2-\alpha^2\epsilon^2 + \sqrt{(r^2-\alpha^2\epsilon^2)^2+4\alpha^2\epsilon^2 r^2 \cos^2\theta}\right)
\label{s2}
\ee

We are interested in the behavior of the Newtonian potential at large distances $r$. We expand $s(r,\theta)$ in \ref{s2} for large $r$ and use the first two of \ref{conversion} and the definition of $\sigma$, to obtain
\be
\sigma \simeq \frac{r}{\alpha\epsilon} \left(1 - \frac{\alpha^2\epsilon^2\sin^2\theta}{2 r^2} + {\mathcal O}(r^{-4}) \right), \, \;\;\; \cos\xi\simeq \cos\theta +{\cal O}(r^{-2}) 
\ee
and from these
\be
\cot^{-1}\sigma \simeq \frac{1}{\sigma} - \frac{1}{3\sigma^3} +{\cal O}(\sigma^{-5}) \simeq \frac{\alpha \epsilon}{r} + \frac{\alpha^3\epsilon^3}{2 r^3} \left(\sin^2\theta - \frac{2}{3} \right) + {\cal O}(r^{-5}) \;\;{\rm and} \;\; P_2(\cos\xi) \simeq P_2(\cos\theta) +{\cal O}(r^{-2})
\ee
Substituting, finally, the above into the Newtonian potential \ref{VN}, we get
\be
V_N \simeq - \frac{M}{r}  \left(1 - \frac{\alpha^2 \epsilon^2}{5 r^2} P_2(\cos\theta) \right) + {\mathcal O}(r^{-5})
\ee 
Adding to $2V_N$ the repulsive potential $-H_0^2 r^2$ representing the effect of the cosmological constant at large distances from the oblate ellipsoidal body, we obtain in the Newtonian approximation of General Relativity the effective gravitational potential at $(r,\theta,\phi)$
\be
U_{\rm eff} \simeq - \frac{2M}{r}  \left(1 - \frac{\alpha^2 \epsilon^2}{5 r^2} P_2(\cos\theta) \right) - H_0^2 r^2
\ee
For the spherical body, $\epsilon=0$, we obtain the well known result $R^{(0)}_{\rm TA,max}=(M/H_0^2)^{1/3}$ \cite{PT14}.

The fractional deviation of the effective potential from the one for a spherical structure is of the order of ${\cal O}(\alpha^2/r^2)$. Correspondingly, the change in the maximum turnaround radius, $\delta R_{\rm TA,max}(\theta)=R^{(\epsilon)}_{\rm TA,max}(\theta) - R^{(0)}_{\rm TA,max}$, for fixed $\theta=\theta_0$, as obtained by the condition $U'=0$, is 
\be
\frac{\delta R_{\rm TA,max}(\theta_0)}{R^{(0)}_{\rm TA,max}} \simeq -\frac{\epsilon^2}{5} \left(\frac{\alpha}{R^{(0)}_{\rm TA,max}}\right)^2 P_2(\cos\theta_0)
\label{deltaRnewton}
\ee
whose average over the solid angle $(\theta_0,\phi)$ is zero, $\left\langle\delta R_{\rm TA, max}(\theta_0)/{R^{(0)}_{\rm TA, max}} \right\rangle \simeq  0$. 

For $\theta_0=0, \pi$ and $\theta_0=\pi/2$, in particular, we obtain
\be
\frac{\delta R_{\rm TA,max}(0, \pi)}{R^{(0)}_{\rm TA,max}} \simeq -\frac{\epsilon^2}{5} \left(\frac{\alpha}{R^{(0)}_{\rm TA,max}}\right)^2 \;\;\;\;\; {\rm and} \;\;\;\;\; 
\frac{\delta R_{\rm TA,max}(\pi/2)}{R^{(0)}_{\rm TA,max}} \simeq \frac{\epsilon^2}{10} \left(\frac{\alpha}{R^{(0)}_{\rm TA,max}}\right)^2
\label{deltaR/R}
\ee
respectively. The result has the qualitative features expected on the basis of Newtonian gravity. The $R_{\rm TA,max}$ becomes smaller (larger) when the attraction to the origin diminishes (increases). In the case of a pancake-like axisymmetric structure the gravitational attraction on the symmetry axis, $\theta_0=0$ (the equatorial plane, $\theta_0=\pi/2$) is weaker (stronger) than it would be if all its mass were at its center. Note that even for $\epsilon\sim {\cal O}(1)$, far away from spherical symmetry, these are very small for realistic structures. Also, on the basis of Newtonian gravity the reader can easily convince her/himself that in the case of a prolate (cigar-shaped) spheroidal structure the behaviour of $\delta R_{\rm TA,max}$ as a function of $\theta_0$ is opposite to the one in formulae  \ref{deltaRnewton} and \ref{deltaR/R}. In particular, it is positive on the symmetry axis $\theta_0=0,\pi$ and negative on the plane $\theta_0=\pi/2$. 

Thus, as long as the mass of a static axisymmetric structure is basically within a radius much smaller than its turnaround size, the dependence of its maximum turnaround radius on its detailed shape is very much suppressed.

\section{A static axisymmetric metric}\label{non-sph}
Having discussed the Newtonian toy model, let us consider the general relativistic cases.
 We shall first consider static intrinsically axisymmetric structure modelled by the metric studied in \cite{Glampedakis:2005cf, Johannsen:2010xs}. In those articles (see also~\cite{Loeb:2013lfa} and references therein), a rotating axisymmetric metric (with $H_0=0$) was introduced to study the uniqueness properties of a stationary black hole, given by,
\begin{equation}
\begin{aligned}
g_{tt} 
&= -\left(\frac{{\widetilde\Delta_r} -a^2 \sin^2\theta}{\rho^2}\right) + \frac{10\, \tilde\epsilon P_2(\cos\theta)}{16 M^2r^2}
\bigg[ 2M \left( 3r^3-9Mr^2+4M^2r+2M^3 \right) - 3r^2(r-2M)^2 \ln\left( \frac{r}{r-2M} \right) \bigg],\\
g_{t\phi} 
&= -\frac{2Mar\sin^2\theta}{\rho^2},\\
g_{rr} 
&= \frac{\rho^2}{\widetilde\Delta_r} + \frac{10\, \tilde\epsilon P_2(\cos\theta)}{16M^2(r-2M)^2}  \bigg[ 2M\left( 3r^3-9Mr^2+4M^2r+2M^3 \right) -3r^2(r-2M)^2 \ln\left( \frac{r}{r-2M} \right) \bigg],\\
g_{\theta \theta} 
&= \rho^2 + \frac{ 10\, \tilde\epsilon \,r P_2(\cos\theta)}{16 M^2} \bigg[2M \left(3r^2+3Mr-2M^2\right) 
 - 3r\left(r^2-2M^2\right) \ln\left( \frac{r}{r-2M} \right) \bigg],\\
g_{\phi \phi} 
&= \left[r^2+a^2+\frac{2Ma^2r\sin^2\theta}{\rho^2} + \frac{ 10\, \tilde\epsilon\, r P_2(\cos\theta)}{16 M^2} 
 \left[2M \left(3r^2+3Mr-2M^2\right) - 3r\left(r^2-2M^2\right) \ln\left( \frac{r}{r-2M} \right) \right] \right]\sin^2\theta,
\end{aligned}
\label{QKmetric}
\end{equation}
where $\tilde\epsilon$ is a dimensionless  parameter and
\begin{equation}
{\widetilde\Delta_r} =r^2-2Mr+a^2
\quad\text{and}\quad
\rho^2= r^2+a^2\cos^2 \theta.
\label{deltasigma}
\end{equation}
Setting ${\tilde\epsilon} \to 0$ one recovers the Kerr spacetime written  in the Boyer-Lindquist coordinates. The above metric usually possesses a naked curvature singularity rendering it unphysical globally. Thus one needs to impose  a cut off radius inside which the metric gets replaced with a suitable interior one.

We shall be concerned about the {\it static} (i.e., non-rotating) limit of the above metric, $a=0$, and use it to model the gravitational field of a large scale structure. The parameter ${\tilde\epsilon}$ thus represents some intrinsic non-sphericity in the shape of the structure, analogous to the eccentricity parameter $\epsilon$ in the previous section. Permissible solutions for $a=0$ require ${\tilde\epsilon} \geq-0.8 $~\cite{Loeb:2013lfa}. Setting ${\tilde\epsilon}=0$ further reduces the metric to the Schwarzschild. 

We are interested in structures with $M\ll R_{\rm TA, max} \ll H_0^{-1}$, i.e. much bigger than their Schwarzschild radius $2M$ and much smaller than the size of the observed Universe, so that they fit inside it. Accordingly, we shall be interested in the weak field regime of \ref{QKmetric}, appropriate to the study of the turnaround radius of such structures, including the leading modification due to a positive $\Lambda$.
 
Expanding \ref{QKmetric} (with $a=0$) up to ${\cal O}(M^3/r^3)$, appropriate to study the weak gravity regime,  we obtain
\begin{eqnarray}
&&g_{tt} \approx -\left(1-\frac{2M}{r}  +  2 \tilde\epsilon P_2(\cos\theta)\frac{M^3}{r^3}  \right)\,, \quad g_{rr} \approx \left(1-\frac{2M}{r} \right)^{-1} - 2 \tilde\epsilon P_2(\cos\theta)\frac{M^3}{r^3} \nonumber\\
&&g_{\theta \theta} \approx r^2 \left(1  - 10 \tilde\epsilon P_2(\cos\theta) \frac{ M }{r}  + 16 \tilde\epsilon P_2(\cos\theta)\frac{ M^3 }{r^3}\right) \, , \;\;\;\; g_{\phi\phi} = g_{\theta\theta} \sin^2\theta 
\end{eqnarray}
From the above weak field expansion it is clear that, $\tilde \epsilon$ cannot be arbitrarily large, for otherwise it would indicate violation of Newton's law. 
The leading modification of the above metric functions in the presence of a positive $\Lambda$
will be to make the replacement,
$$1-\frac{2M}{r}\;\; \longrightarrow \;\; 1-\frac{2M}{r}-H_0^2 r^2 $$
Expanding now the dispersion relation for a test particle following a timelike geodesic, $u_a u^a=-1$, in the above background we obtain,
\begin{eqnarray}
\left(\frac{dr}{d\tau}\right)^2= E^2  - \left(1-\frac{2M}{r}-H_0^2 r^2 + 2 \tilde\epsilon P_2(\cos\theta) \frac{M^3}{r^3}  \right) \left[\frac{L^2}{g_{\phi \phi}} + g_{\theta \theta}\left(\frac{d\theta}{d\tau}\right)^2 +1\right]
\label{geod}
\end{eqnarray}
where $\tau$ is the proper time along the trajectory. We have also defined, owing to the time translation and azimuthal symmetries of the spacetime, the conserved energy and the orbital angular momentum of the test particle,
$$E= - g_{ab} (\partial_t)^a u^b= -g_{tt} \frac{dt}{d\tau} \,, \qquad L=  g_{ab} (\partial_{\phi})^a u^b = g_{\phi \phi}\frac{d\phi}{d\tau} $$
The maximum turnaround condition is obtained by setting $d^2 r/d\tau^2=0$ in \ref{geod}. Note that there can be analogous turnaround condition in the polar direction $\theta$, as well. However, the surface of the compact axisymmetric structure we are looking into is spanned by $\theta$ and $\phi$. Hence any turnaround condition in the polar direction will not carry any information about the maximum size of the structure. Accordingly, for our current purpose we shall only be concerned about the turnaround condition along the radial direction.

The second term on the right hand side of \ref{geod} can be interpreted as the effective potential ($r, \theta$ and velocity dependent), the test particle experiences in the static and axisymmetric gravitational field of the structure. It is a positive definite quantity in our region of interest.  Let us now imagine a test particle approaching the maximum turnaround point, where the effective potential has a maximum and the radial speed becomes, by definition, zero or vanishingly small. Sufficiently close to that point, the potential must be monotonically increasing. Now since the quantity appearing in the square bracket is greater than or equal to unity, it is clear that the maximum upper bound of all the turnaround radii, i.e. $R_{\rm TA,max}$, will simply correspond to $L=0= d\theta/d\tau$ in \ref{geod}. Equivalently, any motion along the angular directions will always create centrifugal force on the test particle at least at the leading order, which will  reduce the turnaround radius. 

Thus   $R_{\rm TA, max}$ is found by setting the first radial derivative of the resulting effective potential to zero, keeping the angle $\theta=\theta_0$ as an input parameter. 
We find the leading correction $\delta R_{\rm TA,max}$ over the spherically symmetric case due to the non-sphericity parameter $\tilde{\epsilon}$, 
\begin{eqnarray}
\frac{\delta R_{\rm TA, max}(\theta_0)}{R^{(0)}_{\rm TA, max}} = -\tilde{\epsilon} \, P_2(\cos \theta_0)\left( \frac{M}{R^{(0)}_{\rm TA, max}}\right)^2
\label{RT}
\end{eqnarray}
whose average over all directions $(\theta_0,\phi)$ vanishes. Note in particular that if we take $\tilde{\epsilon}$ to be positive, the above result is in qualitative  agreement with that of the previous section, thereby describing a pancake shaped  structure. On the other hand, for $\tilde{\epsilon} <0$, (\ref{RT}) corresponds to a prolate  ellipsoidal structure. Note also that \ref{deltaR/R} contains the actual length scale of the structure, $\alpha$, whereas the above formula contains its Schwarzschild radius. Accordingly, we expect \ref{RT} would be small compared to \ref{deltaR/R}, for typical structures.

\section{The case of the Kerr-de Sitter spacetime}\label{kds}
We shall next study the example of the Kerr-de Sitter spacetime. This is meant to be a simple model of a structure, which is not inherently non-spherical -- but whose departure from spherical symmetry is due to its angular momentum. Accordingly, we expect to find qualitatively different features in $R_{\rm TA, max}$, compared to what we have seen so far.  

The Kerr-de Sitter spacetime represents an axisymmetric  structure, spinning along an azimuthal direction with respect to a given axis, embedded in the de Sitter universe. The spacetime is  axisymmetric and stationary.  It asymptotically approaches the de Sitter spacetime at  large radial coordinate. Since the de Sitter spacetime is spherically symmetric and also static inside the cosmological event horizon, it is natural to choose a coordinate system of the Kerr-de Sitter spacetime such that (a) it is manifestly stationary and axisymmetric and (b) it asymptotically coincides with that of the de Sitter.  Such a description is realized by the so called Boyer-Lindquist coordinates~\cite{Carter:1968ks}, 
\begin{eqnarray}
ds^2=-\frac{\Delta_r-a^2\sin^2\theta \Delta_{\theta}}{\rho^2}dt^2 -\frac{2a \sin^2 \theta }{\rho^2 \Xi} \left( (r^2+a^2 )\Delta_{\theta}-\Delta_r\right)dt d\phi \nonumber\\ + \frac{\sin^2 \theta }{\rho^2 \Xi^2} \left( (r^2 +a^2)^2 \Delta_{\theta}-\Delta_r a^2 \sin^2\theta\right)d\phi^2 + \frac{\rho^2}{\Delta_r}dr^2 + \frac{\rho^2}{\Delta_{\theta}}d\theta^2
\label{sup1}
\end{eqnarray}
where,
\begin{equation}
\Delta_r = (r^2 +a^2) \left(1-H_0^2 r^2 \right) -2Mr, \quad \Delta_{\theta} =1 + H_0^2 a^2 \cos^2\theta,  \quad \Xi = 1+ H_0^2 a^2,  \quad \rho^2 =r^2 +a^2 \cos^2 \theta\,, 
\end{equation}
 $M$ is the mass parameter of the structure and $a=J/M$ is its angular momentum per unit mass. For $a=0$, in particular, we recover the Schwarzschild-de Sitter spacetime, whereas setting in addition $M=0$ it reduces to the de Sitter spacetime written in the static patch.  

The metric \ref{sup1} is $t-$ and $\phi-$ independent. The corresponding conserved energy and angular momentum of the test particle with velocity $u^a=dx^a/d\tau$, where $\tau$ is the proper time along the trajectory, are $E = -g_{tb} u^b$ and $L = g_{\phi b} u^b$. These, along with the  expansion of the dispersion relation ($u_au^a=-1$) for a timelike geodesic in the background of \ref{sup1} gives~\cite{Carter:1968ks},
\begin{eqnarray}
&&\frac{dt}{d\tau}={\frac{a \Xi}{\Delta_r \Delta_{\theta} \rho^2} \left[\Delta_r\left(L-\frac{Ea\sin^2\theta }{\Xi} \right)- \Delta_{\theta} (r^2+a^2) \left(L - \frac{E(r^2+a^2)}{a\Xi} \right) \right] }    \nonumber \\ 
&&\frac{d\phi}{d\tau}={ \frac{\Xi^2}{\Delta_r \Delta_{\theta }\rho^2 \sin^2\theta } \left[\Delta_r\left(L-\frac{Ea\sin^2\theta}{\Xi} \right) -a^2\sin^2\theta \Delta_{\theta}\left(L-\frac{E(r^2+a^2)}{a\Xi} \right) \right]}\nonumber \\
&&\rho^4\left(\frac{dr}{d\tau}\right)^2 =   { a^2 \Xi^2\left(L-\frac{E(r^2+a^2)}{a\Xi}\right)^2-\Delta_r \left(K_C+r^2\right),}    \nonumber\\ 
&&\rho^4\left(\frac{d\theta}{d\tau}\right)^2 = { -\frac{\Xi^2}{\sin^2\theta}\left(L-\frac{Ea\sin^2\theta}{\Xi} \right)^2 +\Delta_{\theta} \left(K_C- a^2 \cos^2 \theta\right)} \, \equiv \lambda(\theta) 
\label{sup1'}
\end{eqnarray}
where $K_C$ is Carter's constant of variable separation  and $\lambda(\theta)$ is an abbreviation of the right hand side of \ref{sup1'}.  The general expression for the  maximum turnaround radius or zero acceleration condition in the radial direction   can be found from the third and fourth of \ref{sup1'}, by setting as earlier,  $d^2 r/d\tau^2=0  $,
and treating $E$, $L$, $M$, $a$, $K_C$ and $\theta$ as numerical inputs. Note from \ref{sup1'} that  there will be  terms linear in $L$, showing that unlike the static and spherically symmetric case $(a=0)$, the direction of rotation of orbits will be distinguished here.  In particular, using $K_C\geq 0$ e.g.~\cite{Bhattacharya:2017scw}, we can show that for the retrograde ($L<0$) orbits, the turnaround radius will be higher than that of the prograde  ($L>0$) ones. 

Note also that just like the case of the static axisymmetric spacetime discussed in the previous section, we shall not consider any turnaround condition along the $\theta$ direction.

As earlier, we shall focus on cosmic structures satisfying the conditions $M \ll R_{\rm TA, max} \ll H_0^{-1}$.  The constraint $a\lesssim {\cal O}(M)$, known in the case of an isolated black hole, does not apply here. Any potential naked singularity is hidden inside the body of the structure, where the above metric is not valid. However, the analysis of simulations \cite{korkidis} shows that the condition $a\lesssim {\cal O}(M)$ is comfortably satisfied by the large scale structures studied here.\footnote{Incidentally, we refer our reader also to~\cite{Zhou:2019kwb} and references therein, for the so called super-spinning Kerr solution, where $a$  can exceed $M$, but at the expense of introducing additional terms in the metric.}

Now, keeping in mind that we shall work essentially in a weak gravity regime, it will be more convenient for us, instead of using \ref{sup1'}, to use a simple alternative derivation of the $R_{\rm TA, max}$ for \ref{sup1} below,  without introducing Carter's constant. We introduce the timelike vector field $\chi^a$, 
$$\chi^a = (\partial_t)^a -\frac{(\partial_t \cdot \partial_{\phi}) }{(\partial_{\phi}\cdot \partial_{\phi})}\,(\partial_{\phi})^a  = (\partial_t)^a -\frac{g_{t\phi} }{g_{\phi \phi}}\,(\partial_{\phi})^a \,$$
It satisfies $\chi\cdot \partial_\phi =0$, while the square of its norm is
$$\chi^a\chi_a =  \frac{g_{tt} g_{\phi\phi} - g^2_{t\phi}}{g_{\phi\phi}} = -\frac{\rho^2 \Delta_r \Delta_\theta}{(r^2+a^2)^2 \Delta_\theta - \Delta_r a^2 \sin^2\theta} $$
which, with $\Delta_r>0$, is easily seen to be negative. In other words, $\chi^a$ is a timelike vector field. 
It is convenient to choose the orthogonal basis for \ref{sup1} : $\{\chi^a,\, (\partial_{\phi})^a,\, (\partial_{\theta})^a,\, (\partial_r)^a\}$. 
 Expanding the dispersion relation, $u\cdot u=-1$  in  this orthogonal basis we obtain
\be
\left(\frac{dr}{d\tau}\right)^2=\frac{\left(E- A(r, \theta)L \right)^2}{\rho^4}\, \left((r^2+a^2)^2\Delta_{\theta} - \Delta_r a^2 \sin^2 \theta\right) -\frac{L^2 \Xi^2\,\Delta_r}{\sin^2 \theta \left((r^2+a^2)^2 \Delta_{\theta}- \Delta_r a^2 \sin^2 \theta \right)}-\frac{\Delta_r }{\rho^2} -\frac{\lambda(\theta) \Delta_r}{\rho^4}
\label{rt1}
\ee
where 
$A(r,\theta)$ is defined by 
$$
A(r, \theta)\equiv \frac{a \Xi\,(2Mr + H_0^2 r^4)}{(r^2+a^2)^2 -\Delta_r a^2 \sin^2 \theta} 
$$
and 
the positive semi-definite $\lambda (\theta)$ is the function appearing on the right hand side of the last of \ref{sup1'}.
It is easy to see then that the leading radial force originating from  this term is repulsive, indicating decrease in the maximum turnaround radius. Thus, in order to find $R_{\rm TA, max}$ we may ignore the kinetic energy of the test particle along the polar angle, compared to that of along the radial  direction. We, thus, set $\theta=\theta_0={\rm constant}$ in \ref{rt1} and obtain
\begin{eqnarray}
\left(\frac{dr}{d\tau}\right)^2=\frac{\left(E- A(r, \theta_0)L \right)^2}{\rho_0^4}\, \left((r^2+a^2)^2\Delta_{\theta_0} - \Delta_r a^2 \sin^2 \theta_0 \right) -\frac{L^2 \Xi^2\,\Delta_r}{\sin^2 \theta_0 \left((r^2+a^2)^2 \Delta_{\theta_0}- \Delta_r a^2 \sin^2 \theta_0 \right)}-\frac{\Delta_r }{\rho_0^2} \nonumber\\
\label{ta3}
\end{eqnarray}
where the subscript $0$ indicates that $\theta$ is replaced with $\theta_0$. 
Notice, although not surprisingly, the above equation indicates that a test particle with $L\neq 0$ cannot be sitting at the poles, $\theta = 0, \pi$. 

Expanding next the right hand side of \ref{ta3} up to the third order of the metric functions we obtain
\begin{eqnarray}
\left(\frac{dr}{d\tau}\right)^2\!\!\!&\approx&\!\!\! E^2 \left[1+\frac{a^2 \sin^2 \theta_0}{r^2} \left(1+\frac{2M}{r}+H_0^2 r^2\right)\right]     -2ELa\left(\frac{2M}{r^3}+H_0^2 \right)\,\,\nonumber\\
&&-\frac{L^2}{r^2 \sin^2 \theta_0}\left[1 -H_0^2 r^2 -\frac{2M}{r}-\frac{a^2 \cos^2 \theta_0}{r^2}-\frac{2Ma^2 \sin^2\theta_0}{r^3}-2H_0^2 a^2 \right] \nonumber\\
&& - \left(1-\frac{2M}{r}-H_0^2 r^2 +\frac{a^2 \sin^2 \theta_0}{r^2} +\frac{2Ma^2 \cos^2\theta_0}{r^3}-H_0^2 a^2\sin^2\theta_0\right)
\label{ta4}
\end{eqnarray}
As a consistency check, we note that as $a \to 0$, the largest root of $d^2r/d\tau^2=0$ corresponds to $L=0$, recovering the result of the static spherically symmetric case, $R^{(0)}_{\rm TA, max} = ({M}/{H_0^2})^{1/3}$~\cite{PT14}.

The distinction between the $L>0$ and $L<0$ trajectories is now apparent.  In particular for $L<0$, the term  proportional to $LaE$ on the right hand side  generates an attractive potential. Note also that the leading of the terms containing $L^2$ creates repulsion, indicating decrease in the size of the maximum turnaround radius. We would thus like to investigate the effect of the interplay between these two terms on $R_{\rm TA, max}$. However, as we argue next these $L-$dependent terms are subleading and can be ignored.

As long as $E\sim {\cal O}(1)$, which is the case of interest to us, \ref{ta4} implies that $L$ may have significant contribution to $R_{\rm TA, max}$ only if (a) it is negative generating an attractive force, {\it and} (b) it is on the order of $L \sim  {\cal O} (R_{\rm TA, max})$, for only then the term linear in $L$, can be expected to become comparable to the other terms. Note also that according to \ref{ta4} the attractive terms proportional to $L^2$ are subleading compared to the dominant repulsive one. 

However, for such high values of $L$, we have in the turnaround region,
$$\frac{L^2}{r^2 \sin^2 \theta_0}\, \gtrsim \, {\cal O}(1)$$
whereas, recalling that $ M \ll R_{\rm TA, max} \ll H_0^{-1}$ we have
$$L a E\left(\frac{2M}{r^3}+H_0^2 \right) \ll 1$$
Thus, the repulsive term of the orbital angular momentum still dominates over the attractive term linear in $L$.  Accordingly, we conclude that even though a negative $L$ generates an attractive force in the Kerr-de Sitter geometry, the $R_{\rm TA,max}$ would still correspond to $L = 0$. Thus, the leading shift $\delta R_{\rm TA,max}(\theta_0)$ of $R^{(0)}_{\rm TA, max}$ is given by
\begin{eqnarray}
\frac{\delta R_{\rm TA, max}(\theta_0)}{R^{(0)}_{\rm TA, max}} &\simeq &  \frac{a^2( E^2 \sin^2 \theta_0 - \cos^2 \theta_0)}{({R^{(0)}_{\rm TA, max})^2}}  + \frac{a^2 \, \sin^2 \theta_0\,(E^2-1)}{3M R^{(0)}_{\rm TA, max}} \nonumber \\
&\simeq & - \left(\frac{a}{R^{(0)}_{\rm TA, max}} \right)^2 \left((1-3H_0^2 R^{(0) 2}_{\rm TA,max})\cos^2\theta_0 + 3H_0^2 R^{(0) 2}_{\rm TA,max} \right) 
\label{ta4new2}
\end{eqnarray}
where in the last step we used for the energy $E$ its value for a particle at rest in the maximum turnaround region
$$
E= - g_{tt} \frac{dt}{d\tau} = \sqrt {-g_{tt}}\simeq 1-\frac32 H_0^2 \left(R^{(0)}_{\rm TA, max}\right)^2 
$$
which is consistent with our assumption $E\simeq 1$. The average of \ref{ta4new2} over the solid angle $(\theta_0,\phi)$ is $\left\langle\delta R_{\rm TA, max}(\theta_0)/{R^{(0)}_{\rm TA, max}} \right\rangle \simeq  -a^2/\left(3R^{(0) 2}_{\rm TA,max}\right)$. 

A rough estimate of this quantity for the case of a ``rotating galaxy cluster" can be obtained in the case of small  fractional change of $R_{\rm TA,max}$. Then, with $I, R$ and $M$ the moment of inertia, the size and the mass, respectively, of the cluster, we obtain $a=J/M \sim I\omega / M \sim R^2 \omega \lesssim R \,v_{\rm out}$, with $v_{\rm out}$ the azimuthal speed of the outmost galaxies in the structure. Thus, we have roughly $\left\langle\delta R_{\rm TA, max}(\theta_0)/{R^{(0)}_{\rm TA, max}}\right\rangle \sim {\cal O}\left(v_{\rm out}^2\right)$, which is much smaller than unity in realistic structures.

Note in particular that \ref{ta4new2} shows decrease in the value of $R_{\rm TA, max}$ compared to the case of a spherically symmetric structure on the axis ($\theta_0 =0, \pi$) as well as on the equatorial plane ($\theta_0 = \pi/2$). This is in contrast to the cases discussed in \ref{newton} and \ref{non-sph}, making the Kerr-de Sitter qualitatively different. Such decrease could be understood as the repulsive effect originating from the spacetime rotation. Note also that the decrease is minimum at $\theta =\pi/2$ and it monotonically increases to its maximum value at $\theta=0, \pi$. This could be understood as the flattening of the structure on the equatorial plane due to its rotation.

\section{Conclusions}\label{conc}
As we explained in the Introduction, despite the fact that actual galaxy clusters in the Universe are neither isolated nor spherically symmetric, the study of the turnaround radius of such a structure in the context of $\Lambda$CDM, provided useful information and quantitative agreement with the measured values of the sizes of realistic superclusters. Although this already indicates that the assumption of spherical symmetry and of no influence of neighbouring structures is a good approximation, direct quantitative study was still missing.  Accordingly as a first step which can be treated analytically and hence provides some intuitive understanding, we focused on axisymmetric structures, whose departure from spherical symmetry is due either to intrinsic non-sphericity or to rotation.

As we have discussed at the end of \ref{s1}, there could be various ways to generate axisymmetric spacetime geometry and hence requires suitable modelling.  Accordingly, we used three different models to describe such  hypothetical idealised structures. A homogeneous ellipsoid in background de Sitter space treated in the Newtonian approximation, a static axisymmetric spacetime modelling an intrinsically asymmetric structure and, finally the Kerr-de Sitter spacetime appropriate to describe a stationary axisymmetric background, respectively in \ref{newton}, \ref{non-sph} and \ref{kds}. Note that the last one is not intrinsically axisymmetric, but is so due to the rotation of the structure.

Our main conclusion is that in all these cases the fractional change $\delta R_{\rm TA, max}(\theta_0)/{R^{(0)}_{\rm TA, max}}$ of the maximum turnaround size of a structure due to the departure from spherical symmetry is proportional to the product of the corresponding eccentricity parameter times the square of a small ratio either of two characteristic length scales (\ref{newton} and \ref{non-sph}) or of the azimuthal speed of the outmost galaxies over the speed of light (\ref{kds}). Thus, $\delta R_{\rm TA, max}(\theta_0)/{R^{(0)}_{\rm TA, max}}$ and its angular average $\left\langle\delta R_{\rm TA, max}(\theta_0)/{R^{(0)}_{\rm TA, max}}\right\rangle$ in all models is predicted to be negligible, for practically any value of the relevant eccentricity parameter. 

Furthermore, we showed that in the Kerr-de Sitter case, analysed in \ref{kds}, the change $\delta R_{\rm TA,max}(\theta_0)$ in the turnaround radius is negative for all values of the polar angle. This is in contrast to the examples studied in \ref{newton} and \ref{non-sph}, concerning structures with intrinsic non-sphericity in their shapes, in which $\delta R_{\rm TA,max}(\theta_0)$ has positive and negative values, while it vanishes when averaged over the angles. Hence it shows a qualitative difference between an intrinsically axisymmetric structure and a rotating structure. 

Thus for the geometries we have considered, we conclude that the fractional change of the maximum turnaround size of a bound static or stationary axially symmetric structure characterised by $M \ll R_{\rm TA,max} \ll H_0^{-1}$ is negligible. It is also evident from our analysis that  this conclusion holds for all realistic values of the parameters specifying axisymmetry, because of the presence of other large suppression factors.

 Based on this result and on the fact that  realistic galaxy clusters or superclusters in the universe typically look more spherical than cigar- or pancake-shaped structures discussed here, we would like to speculate that the turnaround radii of realistic cosmic structures also depend very little on their shapes. In fact, this has been confirmed by a phenomenological analysis of the turnaround radii of  large galaxy clusters obtained in extensive $N$-body numerical simulations \cite{korkidis}. This general analysis shows that like the axisymmetric structures we have considered here, the turnaround radius depends very little on their shape. This is because typically the turnaround scale is located far away from the bulk of the mass of a structure and  the higher order multipoles of the gravitational potential are sub-leading at this length scale. In the analytical front however, a more detailed quantitative treatment of the general shape dependence of the turnaround size of cosmic structures is missing and certainly this warrants further investigation.

\section*{Acknowledgements}
SB's research is partially supported by the ISIRD grant 9-289/2017/IITRPR/704. TNT wishes to thank G. Korkidis and especially V.~Pavlidou for many insightful discussions and valuable comments.


\end{document}